\documentclass[useAMS,usenatbib]{mn2e}
\usepackage{graphicx}
\newcommand\apjl{{ApJL}}%
\newcommand\aj{{AJ}}%
\newcommand\araa{{ARA\&A}}%
\newcommand\apj{{ApJ}}%
\newcommand\apjs{{ApJS}}%
\newcommand\apss{{Ap\&SS}}%
\newcommand\aap{{A\&A}}%
\newcommand\mnras{{MNRAS}}%
\newcommand\nat{{Nature}}%
\newcommand{\lsim}{\mathrel{\hbox{\rlap{\lower.55ex\hbox{$\sim$}} \kern-.3em \raise.4ex \hbox{$<$}}}} 

\title[Submillimetre observations of RX\,J1856.5--3754]{Submillimetre observations of RX\,J1856.5--3754\thanks{Based on ESO observations made with the APEX-telescope at Llano de Chajnantor under programme 81.D-0777}}
\author[B. Posselt et al.]{B. Posselt$^{1}$\thanks{E-mail:
bposselt@cfa.harvard.edu}, 
K. Schreyer$^{2}$, R. Perna$^{3}$, M.W. Sommer$^{4}$, B. Klein$^{5}$, P. Slane$^{1}$
\\
$^{1}$ Harvard-Smithsonian Center for Astrophysics, 60 Garden Street, 02138 Cambridge, MA, USA\\
$^{2}$ Astrophysikalisches Institut und Universit\"{a}ts-Sternwarte Jena, Schillerg\"{a}sschen 2-3, D-07745 Jena, Germany\\
$^{3}$ 440 UCB, U. of Colorado, Boulder, 80309, USA\\ 
$^{4}$ Argelander-Institut f\"{u}r Astronomie, Auf dem H\"{u}gel 71, D-53121 Bonn, Germany\\
$^{5}$ Max-Planck-Institut f\"{u}r Radioastronomie, Auf dem H\"{u}gel 69,D-53121 Bonn,Germany\\
}
\begin{document}


\pagerange{\pageref{firstpage}--\pageref{lastpage}} \pubyear{2010}

\maketitle

\label{firstpage}

\begin{abstract}
We report on submillimetre bolometer observations of the isolated
neutron star RX\,J1856.5--3754 using the LABOCA bolometer array on the
Atacama Pathfinder Experiment (APEX) Telescope.  No cold dust
continuum emission peak at the position of RX\,J1856.5--3754 was
detected. The $3 \sigma$ flux density upper limit of 5\,mJy translates
into a cold dust mass limit of a few earth masses.  We use the new
submillimetre limit, together with a previously obtained H-band limit,
to constrain the presence of a gaseous, circumpulsar  disc.
Adopting a simple irradiated-disc model, we obtain a mass accretion
limit of $ \dot{M} \lsim 10^{14}$ g s$^{-1}$, and a maximum outer disc
radius of $\sim 10^{14}$\,cm.  By examining the projected proper
motion of RX\,J1856.5--3754, we speculate about a possible encounter
of the neutron star with a dense fragment of the CrA molecular cloud a
few thousand years ago.
\end{abstract}

\begin{keywords}
submillimetre- stars: neutron- X-rays: individual: RX\,J1856.5--3754
\end{keywords}

\section{Introduction}
Since the discovery of planets around the pulsar PSR~1257+12 by
\citet{WF92}, dusty discs around pulsars have become interesting to
observers, who have been trying to detect them in the infrared (IR) or at
(sub-)millimeter wavelengths.  Most comprehensive are the searches by
\citet{loehmer2004} and \citet{Greaves2000}. Previous searches had
concentrated mainly on recycled, thus formerly accreting, radio
pulsars as the most likely objects to be surrounded by dusty
discs. Neither of the above mentioned surveys detected dust emission
around the planet-hosting PSR~1257+12, nor did the recent search with
$Spitzer$ at 24\,$\mu$m and at 70\,$\mu$m by \citet{Bryden2006}.
However, dusty discs could also be present around non-recycled neutron
stars (NSs). Following the supernova explosion, which creates the NS,
some of the explosion ejecta may fail to escape and remain bound -
forming a fallback disc. Such fallback discs are a general prediction
of current supernova models \citep{Michel1981,Chevalier1989}, and
have been invoked by a number of authors to explain a variety of
phenomena related to NSs (e.g., \citealt{Chatterjee2000,Alpar2001,Menou2001,Blackman2004}).
Recently, \citet{Wang2006} reported the
discovery of mid-infrared emission from a cool disc around the Anomalous X-ray
pulsar AXP 4U~0142+61. 
\citet{Wang2006} interpreted their detection as a passive disc, while
\citet{Ertan2007} argued that it could also originate from an actively accreting disc.
To date, AXP 4U~0142+61 is the only isolated neutron star for which a
fallback disc is believed to have been detected.  Such discs appear to
be rare. According to \citet{Eksi2005} fallback disc are rare because
they are likely to be disrupted when the newly born NS spins rapidly
through the propeller stage, at which in-flowing matter, instead of being
accreted, would be expelled. The fallback discs can survive if the initial NS spin
is slow enough ($\geq 40$~ms at a magnetic moment of $\mu=10^{30}$~G
cm$^{3}$).
\citet{Jones2007} studied the effect of pulsar wind induced ablation of fallback discs. He concluded  that long-lived discs could be present in
many pulsars without exceeding published limits on IR luminosity.\\ 
In the following, we report on submillimetre
observations of RX\,J1856.5--3754, which is the brightest and closest
member of a class of NSs neglected so far in the search for
circumstellar discs, the X-ray thermal isolated neutron stars. They
are peculiar because they show pure thermal soft X-ray spectra without
any (confirmed) non-thermal emission, especially no confirmed radio
emission; for reviews, see \citet{Kaplan2008A} or \citet{Haberl2007}.
These objects have periods in the range of 4 to 12~s. Thus, they are
much slower than the bulk of radio pulsars. Their X-ray pulse periods  and period
derivatives are similar to those of the AXPs and soft
gamma-ray repeaters, SGRs, (see, e.g., \citealt{Kaplan2009}).  This has led to discussions about whether they may be
related to those objects.  For example, \citet{Alpar2001, Alpar2007}
suggested that the X-ray thermal isolated NSs may simply have accretion
discs with smaller masses than those of the AXPs. 
We note that it is currently not clear whether AXPs host accretion discs. 
The prevailing model for AXPs and SGRs is the magnetar model -- isolated, young neutron stars with exceptionally high ($\approx 10^{14}$\,G) surface magnetic fields \citep{Duncan1992}.\\

RX\,J1856.5--3754 has been observed at radio
wavelengths (to date without detection), near infrared (without detection),
optical and X-ray wavelengths. 
The striking absence of deviations from a pure blackbody in its X-ray spectrum has
led to lively discussions on the nature of the object and of its X-ray emission,  
including highly magnetized atmospheres (e.g.,
\citealt{Ho2007a}), condensed surfaces (e.g., \citealt{Burwitz2003}) and quark
stars (e.g., \citealt{Drake2002}). 
The most recent, preliminary, parallax by \citet{Kaplan2007a} is $167^{+18}_{-15}$\,pc and
marks RX\,J1856.5--3754 as one of the closest NSs currently known.
\citet{Tiengo2007} discovered the $\approx 7$\,s pulsations of RX\,J1856.5--3754,
which has an extremely small  pulsed fraction in the 0.15-1.2 keV range of only 1.2\,\%. 
A timing study by \citet{Kerkwijk2008} inferred a magnetic field of $1.5 \times
10^{13}$\,G assuming spin-down by dipole radiation. 
Interestingly, the spin-down luminosity is not high enough to explain the  H{$\alpha$} nebula around
RX\,J1856.5--3754, discovered by \citet{Kerkwijk2001}.  The nature of the
H{$\alpha$} nebula remains unclear to date (see, e.g., \citealt{Kerkwijk2008}).
These authors also derived a characteristic age of $~4$\,Myrs, which
is much older than the kinematic age, $~0.5$\,Myrs, obtained assuming
an origin in the Upper Scorpius OB association (e.g.,
\citealt{Walter2002}). At such ages, any fallback disc around the
source would be expected to be rather cool.\\

\begin{figure}
\includegraphics[width=8.5cm]{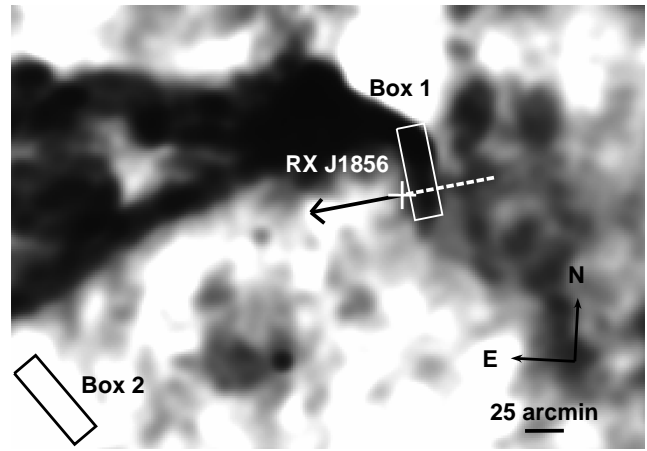}
 \caption{The IRAS/IRIS $100\mu$m band of the neighbourhood of
 RX\,J1856.5--3754.  Strong infrared emission, signaling the presence of dust,
is  shown by darker areas. The peak dust emission outlines the CrA star
 forming region. The position of RX\,J1856.5--3754, as well the
 direction of its proper motion, is indicated. The dashed white line
 shows the rough projection of the proper motion into the past. About
 3000 years ago the neutron star might have crossed a region of the
 denser dust filament -- depending on the actual distances of the
 filament and RX\,J1856.5--3754.  The boxes 1 and 2 cover the same
 area, see section\,\ref{past} for discussion. }
\label{fig:IRAS}
\end{figure}

Within distance error bars, $117 \pm 12$\,pc by \citet{Walter2002} to $167^{+18}_{-15}$\,pc by \citet{Kaplan2007a}, 
RX\,J1856.5--3754 appears to
be in the outskirts of the Corona Australis star-forming cloud, whose
distance is relatively well known: $129 \pm 11$\,pc from the orbit solution of the double-lined spectroscopic binary TY CrA by \citet{Casey1998}; for a detailed discussion we refer to Sect.\,1.3 in \citet{Neuh2008}.  
Bondi-Hoyle accretion \citep{Bondi1944,Bondi1952} 
from the interstellar medium (ISM) is unlikely to be a major contributor to the observed
X-ray luminosity today, given the high velocity of this neutron star
\citep{Kerkwijk2001}.  Furthermore, it is likely that isolated NSs
accrete at sub-Bondi rates \citep{Perna2003}. 
The accretion rate scales inversely with the magnetic moment as $~{\mu}^{-2.1}$ according to \citet{Toropina2006, Romanova2003}. RX\,J1856.5--3754 has a high magnetic field, $1.5 \times 10^{13}$\,G \citep{Kerkwijk2008}, reducing the possibility of accretion even further.
However, as \citet{Drake2003} noted,
RX\,J1856.5--3754 might have passed more dense regions than it now
resides in. Indeed the IRAS 100\,$\mu$m image shows some cloud
fragments projected along the past trajectory of RX\,J1856.5--3754 (see
Fig.~\ref{fig:IRAS}), opening the possibility that the neutron star
collected material in the past.\\ 
Here, we investigate the surroundings of RX\,J1856.5--3754 for cold
dust as one might expect either in an old fallback disc or in a
dense gas disc.  
The gas content of a fallback disc is not well known. The formation from supernova ejecta would be expected to produce a disc mainly consisting of heavy elements. 
\citet{Currie2007} showed that gas from viscous, circumpulsar discs typically depletes on
very short, $\approx 10^5$ yr, timescales due to high temperatures during the early evolutionary stages, subsequent rapid spread of the disc and cooling allowing for the heavy elements to condense onto grains.
In addition, \citet{Phillips1994} have shown that for a pulsar moving with $\approx
100$\,km\,s$^{-1}$ through the ISM, small dust grains with radii of $\leq 0.1$\,$\mu$m
are spiraled in from a circumpulsar disc to the neutron star on a time
scale of order $~ 10^6$ years. 
Thus, a circumpulsar disc at the age of RX\,J1856.5--3754 
is likely to be similar to debris discs, which are dust-rich and dominated by larger grains (see, e.g., \citealt{Krivov2009}) as opposed to protostellar discs, which are gas-rich.
If, on the other hand, RX\,J1856.5--3754 has collected new, gas-rich material from the outskirts of the Corona Australis star-forming cloud, the disc composition could 
more closely resemble a protostellar disc.
Given the uncertainties in the gas content and composition of a possible disc around RX\,J1856.5--3754, we will consider here both types of discs -- dense, gas-rich protostellar-type discs, and gas-poor, dust-rich debris-type discs.

\section{Observations and data reduction}
RX\,J1856.5--3754 was observed from the 18th to the 24th of April 2008
using the 295-element Large Apex Bolometer Camera (LABOCA, \citealt{Siringo2009})
on the Atacama Pathfinder Experiment (APEX) telescope \citep{Guesten2006}. 
LABOCA operates at 345GHz (870$\mu$m) with a bandwidth of 60\,GHz. 
The total on-source observing time was 18\,hours. 
Mapping was performed on-the-fly using spiral patterns to ensure even coverage of the source.
To obtain the lowest possible flux limit, the most sensitive part of the array was centered on the source.
The observing conditions were good to excellent with precipitable water vapor levels 
typically below 0.5\,mm. Sky-dips were performed hourly, and combined with 
radiometer readings to obtain accurate opacity estimates as described by \citet{Weiss2008}. 
The absolute flux calibration follows the method outlined by \citet{Siringo2009}, and is expected to be accurate to within $\approx 10\%$ (e.g., \citealt{Greve2009}).
The nominal LABOCA beam is FWHM $18.6 \arcsec \pm 1.0 \arcsec$, and the pointing uncertainty is  $\sim 4\arcsec$\footnote[1]{More on LABOCA calibration
  can be found at:
  http://www.apex-telescope.org/bolometer/laboca/calibration/}.
The data were reduced using the February 2008 release of the BoA reduction
package\footnote[2]{http://www.apex-telescope.org/bolometer/laboca/boa/} (Schuller et al. in prep.).
Reduction steps included the removal of correlated (atmospheric and electronic) noise in several iterations, and the removal of spikes and excessively noisy channels. 

In the second part of the last observation night, we found a sharp drop in the root mean square (rms) of individual bolometer time streams. As these time streams did not contain the expected correlated sky component, these observations (about two hours) were excluded from further consideration. 
To filter out extended emission and residual atmospheric noise at low frequencies, we used direct high-pass filtering of the raw time streams applying the fast Fourier transform (FFT). 
This was done by rejecting frequencies below a chosen cutoff, $\nu_C$.

For each scan, a map with $3^{\prime\prime} \times 3^{\prime\prime}$ sized pixels was constructed, 
weighting the data by the inverse rms of each reduced time stream. 
All maps were then co-added, and the final map was smoothed with the LABOCA beam, $18.6 \arcsec$.
The map rms per beam was computed as the pixel rms in the smoothed map, Fig.~\ref{FigcutFreq0p3}, applying the BoA-task \texttt{computeRms()}.

\section{Results}
\label{results}
\begin{figure}
\includegraphics[width=8.0cm, angle=90]{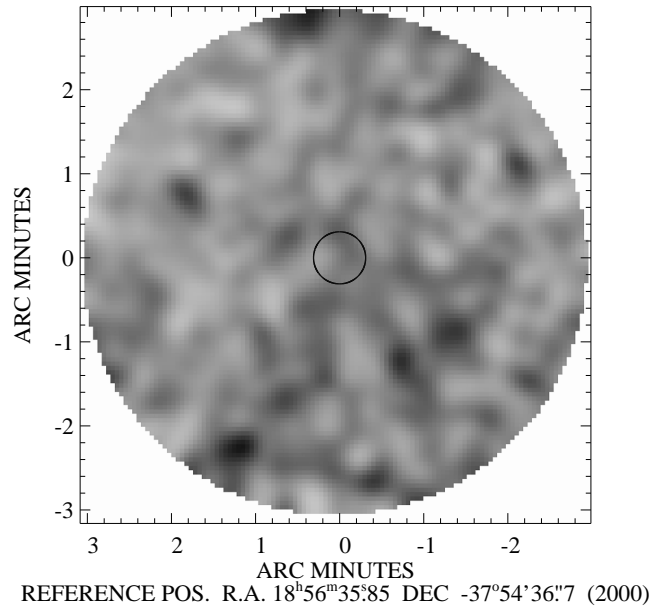}
 \caption{This LABOCA-345\,GHz-continuum-map has been filtered for all modes corresponding 
 to spatial sizes larger than about $150\arcsec$. 
   Darker areas indicate higher submillimetre flux. The circle is
   centered on the position of RX\,J1856.5--3754 with a radius of
   $18.6\arcsec$ (double beam size). No obvious submillimetre point source has been detected at
   the position of RX\,J1856.5--3754. This map has a rms of 1.49 mJy/beam, corresponding to a $3\sigma$ flux density limit of 5 mJy (see text).}
\label{FigcutFreq0p3}
\end{figure}

Fig.~\ref{FigcutFreq0p3} is a filtered image, where
all modes corresponding to spatial sizes larger than about $150\arcsec$ are filtered out 
(the median scan speed is around 96 arcsec s$^{-1}$). 
The same area as visible in Fig.~\ref{FigcutFreq0p3} was considered
when we computed the rms/beam for filtered or unfiltered maps.
No submillimetre-emission has been found at or near the position of
RX\,J1856.5--3754.
The rms has been measured as 3.02 mJy/beam for the map obtained without direct FFT filter, and 1.49
mJy/beam for the direct-FFT-filtered map.
Applying the filter on a test point source of 10 mJy results in recovering of 89\,\% of the original flux.  We correct for this factor and obtain an $3 \sigma$ limit of 5\,mJy/beam for the filtered map of RX\,J1856.5--3754.

\section{Discussion}
\subsection{Dust mass limit}
In contrast to infrared emission, submillimetre emission is nearly always
optically thin and traces the total dust mass.
The conventional formula for calculating this dust mass is:
\begin{equation}
M_d=\frac{F_{\nu} D^2}{B_{\nu}(T_d,\lambda) {{\kappa}^d}_{\nu}(\lambda)}
\end{equation}
where $\nu$ is the frequency, $F_{\nu}$ is the flux density, $D$ is the
distance, $B_{\nu}$ is the
Planck function at a dust temperature $T_d$, 
and $ {{\kappa}^d}_{\nu}$ is the dust mass absorption coefficient. 

The value of $ {{\kappa}^d}_{\nu}$ depends on the grain parameters.
As noted in the introduction, we consider in the following dust
in protostellar discs as well as in debris discs to reflect the wide
range of possible disc compositions around RX\,J1856.5--3754. Since
the gas content in fallback discs is unconstrained, we calculate the
dust mass limits for both disc types. The overall mass of a
protostellar-composition disc would be larger by roughly a factor 100,
the conventional gas-to-dust-ratio. 

In the case of protostellar-disc dust, small grains are present. 
For such protostellar dust, we use $
{{\kappa}^d}_{\nu}=3$\,cm$^{2}$\,g$^{-1}$ according to the relation by \citet{Beckwith1990}: 
${{\kappa}}_{\nu}=0.1 (\nu/10^{12}$\,Hz$)^{\beta}$\,cm$^{2}$\,g$^{-1}$ 
with $\beta=1$, and where ${\kappa}_{\nu}$ is the opacity of dust and gas together.
\citet{Beckwith1990} assumed a gas-to-dust-ratio of 100. 
\citet{Ossenkopf1994} estimated uncertainties for ice-covered
dust grains in protostellar molecular cloud cores to be usually within a factor of 5.
 In the case of debris discs, reported dust mass absorption coefficients at
submillimetre wavelength range from $
{{\kappa}^d}_{\nu}=0.3-1.7$\,cm$^{2}$\,g$^{-1}$
\citep{Dent2000,Najita2005}. 
\\

Dust temperatures, $T_d$, can range from $\approx 10$\,K for cold
interstellar dust to $\approx 1500$\,K (sublimation temperature of
silicate dust). Due to missing constraints in the FIR we assume as a
first approximation that the bulk of the dust visible at $870\mu$m
emits at around 17\,K.  According to the X-ray irradiated disc model
by \citet{Vrtilek1990}, and with the X-ray luminosity of
RX\,J1856.5--3754, $L_X=8.8 \times 10^{31}$\,erg s$^{-1}$ at a distance
of $d=167$\,pc \citep{Drake2002}, this temperature is reached at a
radius of $\approx 7 \times 10^{13}$\,cm from the neutron star.
This value is within the disc sizes discussed in models of
NS fallback discs, e.g., by \citet{Ertan2007}. 
Dust in debris discs is thought to be warmer than the one in protostellar discs, with a range from 10\,K up to a few hundred K, but mostly below 110\,K \citep{Wyatt2008}.
In Table~\ref{massvalues} we give dust mass limits for the
different sets of parameters. 

\begin{table}
 \centering
  \caption{Dust mass limits for the measured $870$\,$\mu$m $3 \sigma$ flux density limit of 5\,mJy at different dust temperatures, $T_d$ and mass absorption coefficients, $ {{\kappa}^d}_{345GHz}$.}
  \begin{tabular}{@{}l c c c@{}}
  \hline
 dust type & $ {{\kappa}^d}_{345GHz}$ & $T_d$ & $M_d$ -limit \\
  &$[$ cm$^{2}$\,g$^{-1}$ $]$ & [K] & [M$_{\earth}$] \\
 \hline
   Protostellar & 3 & 17 & 2.0 \\
   Debris disc & 0.3 & 30 & 9.1 \\
   Debris disc &  1.7 & 17 & 3.6 \\
   Debris disc &  1.7 & 30 & 1.6 \\
   Debris disc &  1.7 & 100 & 0.4 \\
  \hline
  \end{tabular}
 \label{massvalues}
\end{table}

Another uncertainty of the mass estimation is related to the porosity
of the dust.  Porous grains are expected both in the cold
ISM as well as in debris discs. Their influence on $
{{\kappa}^d}_{\nu}$ and $T_d$ can result in masses of a factor three
smaller with respect to masses derived for non-porous grains \citep{2006Vosh}. In case of
fast-moving neutron stars like RX\,J1856, 'sandblasting' by particles
from the ISM fragments the dust \citep{Phillips1994}
and reduces the number of porous grains.

\subsection{Constraints on the presence of a gaseous disc component}

If
discs around isolated NSs are (still) gas-rich, they are
expected to be emitting mostly
in the IR and longer wavelengths.  In the following we use the
 disc model by \citet{Perna2000} to constrain the presence and properties of
a potential gaseous circumstellar disc around RX\,J1856. The model is based on
an active, viscous disc, and takes irradiation through X-rays
into account following the approach by \citet{Vrtilek1990}.  We note
that the model does not include radiative transfer modeling, and 
 it assumes that the disc is optically thick throughout.
However, given our scarce knowledge about
 neutron star discs in general, and the lack of an actual IR or mm detection for RX\,J1856
we restrict to this model in the present work.  

Given the very low X-ray luminosity of RX\,J1856 
($8.8 \times 10^{31}$\,erg s$^{-1}$ at 167\,pc, \citealt{Drake2002}), if there is an
active disc, it must be in the propeller phase, which is
equivalent to saying that the magnetospheric radius, $R_m=6.67\times
\dot{M}/(10^{17} {\rm g}\,{s}^{-1})$ cm, must be smaller than the
corotation radius, $R_{co}=1.5\times 10^8 (P/s)^{2/3}$ cm, 
where $P$ is the period of the X-ray pulsar. Therefore,
in our modelling of the disc emission, the inner radius of the disc,
set equal to $R_m$, is restricted to be smaller than $R_{co}$. With
the magnetic field strength estimate $1.5 \times 10^{13}$\,G by
\citet{Kerkwijk2008}, this yields the condition that only accretion
rate values $\dot{M}\la 10^{16}$ g/s are allowed (or else the star
would be accreting and hence be much brighter in X-rays).

In addition to our submillimetre observations, we use previously
acquired data at near-infrared and infrared wavelengths. 
More specifically,  we use the $H$-band limit derived at
the position of RX\,J1856, $H=21.54 \pm 0.24$\,mag, corresponding to a
flux density of 0.0025\,mJy \citep{Posselt2009}.  
Furthermore, we derive the conservative
IRAS/IRIS flux density limits at the position of the neutron star 
applying the IRIS atlas maps \citep{iris2005,Neugebauer1984}.
These limits,$< 1.9$\,Jy, $< 2.7$\,Jy, $<2.4$\,Jy, and $< 9.6$\,Jy at
12, 25, 60 and 100 microns respectively, are too large to constrain our model. 

\begin{figure}
\includegraphics[width=8.5cm]{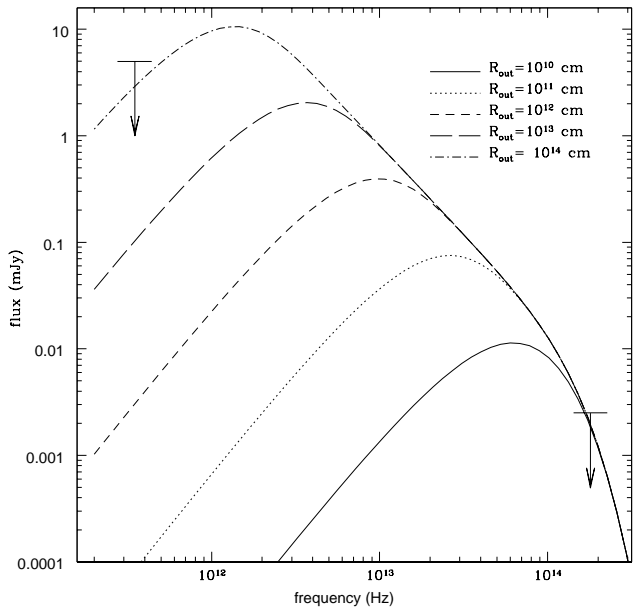}
 \caption{The predicted flux density for an optically thick, gaseous disc around RX\,J1856. 
 The mass accretion is set to $ \dot{M} = 10^{14}$ g s$^{-1}$
(maximum allowed value by the IR limit), and the different curves
show the emission for a range of outer disc radii allowed by the sub-mm
limit. Plotted are also the flux density limits from our APEX
 submillimetre observations and the $H$-band.  }

\label{fig:rosalba}
\end{figure}

Within the allowed range of $\dot{M}$, we then investigated the
expected disc emission as a function of $\dot{M}$, and the disc
boundaries $R_{in}$ and $R_{out}$. We find that $R_{in}$ influences
the emission at wavelengths shorter than the ones considered here, and
therefore the model parameters that can be constrained are $\dot{M}$
and $R_{out}$ (assuming a typical disc inclination of $60^\circ$).  We
find that the H-band limit restricts the accretion rate to be
$\dot{M}\lsim 10^{14}$ g/s, largely independent of $R_{out}$. The
submillimetre $3 \sigma$ limit, on the other hand, restricts the outer
radius to be $R_{out} \lsim 10^{14}$ cm or 7 AU. This outer radius is
essentially independent of the accretion rate since the emission in
the submillimetre is dominated by X-ray
irradiation. Fig.~\ref{fig:rosalba} shows the predicted disc emission
for a range of outer disc radii, and the maximum allowed value of the
accretion rate, $\dot{M}\sim 10^{14}$ g/s.  For the kinematic age of
the pulsar, 0.5 Myrs, and the derived mass accretion limit, we can
estimate the accretion rate $\dot{M}(t_0)$ during the initial time
$t_0\sim 1.5\times 10^{-5}$ $R^{1/2}_{d,7}(t_0)$ yr (Menou et
al. 2001; $R_{d,7}$ is the initial disc radius in units of $10^7$ cm)
before the power law decay begins (Cannizzo, Lee \& Goodman 1990), and
hence the initial disc mass $M(t_0)\sim \dot{M}(t_0)t_0\sim 7 \times
10^{-5} M_{\sun}$.  This upper limit for the initial disc mass is only
a very small fraction of the overall fallback mass discussed to be in
the range of $\sim 0.001M_{\sun}$ to $0.1 M_{\sun}$ by \citet{Lin1991}
and \citet{Chevalier1989}.

\subsection{Looking at the past}
\label{past}
As noted above and shown in Fig.~\ref{fig:IRAS}, the projected trajectory 
of RX\,J1856.5--3754 crossed a CrA molecular cloud fragment roughly
3000 years ago.  The current distance estimates for the CrA cloud
($\approx 130$\,pc; \citealt{Neuh2008}) and
RX\,J1856.5--3754 ($167^{+18}_{-15}$\,pc; preliminary distance by
\citet{Kaplan2007a} or $117 \pm 12$\,pc by \citet{Walter2002}) are
close enough to each other to speculate about a possible encounter
between the neutron star and the cloud fragment.  \citet{Chini2003}
investigated the R Corona Australis molecular cloud in detail by
submillimetre continuum and line observations. Unfortunately, their
map does not cover the cloud fragment west of RX\,J1856. For the cloud
cores that they investigated in the extended eastern region, they
derived densities on the order of $5 \times 10^{-21}$\,g\,cm$^{-3}$ to
$3 \times 10^{-17}$\,g\,cm$^{-3}$. These are a factor $10^3$ to $10^6$
larger than what is usually assumed for the ISM.
If RX\,J1856 crossed such a high density patch,  this could have affected
first a possible 
surrounding fallback disc, destroying it
\citep{Phillips1994} and second, it could have enabled the neutron star
to accrete at higher rates from the ISM.
The extended emission seen in the IRAS
image unfortunately only allows us  a crude estimation of a lower
density limit in the fragment: We consider the IRAS/IRIS 100\,$\mu$m
emission in box 1 centered at the fragment in comparison to the
emission in a box 2 of the same size, but centered around an area devoid
of enhanced IRAS emission (see Fig.~\ref{fig:IRAS}), that supposedly
represents an average density value for the diffuse interstellar
medium. By working with the flux density ratio we can ignore the
problems involved in de-convolving the extended emission from the
instrumental response.  The flux density ratio, translating into a
column density ratio, is 2.6.  The distance to the cloud/fragment is at
least 130\,pc and we take this as the \emph{minimum} distance, $d_2$,
along which the column density, $N_2=n_2 \times d_2$, is derived from
the number density, $n_2$, in box 2.  The angular extent of the
fragment along the direction of the proper motion of RX\,J1856 is
around $18\arcmin$, corresponding to $\approx 1$\,pc at a distance of
130\,pc.  We assume the same extent of the fragment along the
line of sight, $d_1 \approx 1$\,pc;  the column density, $N_1$, in
box 1 is $N_1=n_1 \times d_1 + n_2 \times d_2$.  Therefore, the ratio of
the number density in the fragment to the number density of the
diffuse ISM is at least 200, $n_1/n_2 \geq 200$.  According to
\citet{Phillips1994}, at a number density of $200$\,particles
cm$^{-3}$, the smallest dust grains ($<0.1$\,$\mu$m ) in a fallback disc can be
removed within 3600 years,  assuming a neutron star velocity of
100\,km\,s$^{-1}$ as estimated for RX\,J1856 (e.g.,
\citealt{Kerkwijk2001}). Larger disc particles, however, would still
survive and can be detected in principle in the submillimetre
wavelength regime.  The fact that we do not detect them in our APEX
observation could mean that RX\,J1856 passed a patch of much higher
ISM density, destroying its potential fallback disc.  However, since
the data about a possible encounter between the cloud fragment and
RX\,J1856 is currently insufficient (unknown 3D motion of the neutron
star and distance of the fragment), it is also possible that there
never was a circumstellar disc with a dust mass higher than few earth masses.  

Overall, the above speculations show that a better knowledge of the past neighbourhood of RX\,J1856 is desirable. This can be obtained in principle by further millimeter continuum or molecular line observations. 
 
\section{Conclusions}
We investigated RX\,J1856.5--3754 for $870$\,$\mu$m continuum emission
which would be indicative of a cold dusty disc around the neutron
star.  The derived deep flux density limit translates into a dust mass
limit of few earth masses.  Applying the irradiated (gas-rich)
accretion  disc model by \citet{Perna2000}, together with further
observational constraints, we obtained a mass accretion limit of $
\dot{M} \lsim  10^{14}$ g s$^{-1}$, and a constraint on the outer
disc radius, $R_{out}$, to be smaller than $10^{14}$\,cm or 7\,AU.
Looking at the projected proper motion of RX\,J1856.5--3754, we note
that the neutron star might have passed a dense fragment of the CrA
molecular cloud a few thousand years ago which could have affected a
potential circumstellar disc, as well as have enabled a brief history
of accretion form the ISM.  

\section{acknowledgments}
We thank Axel Weiss for kindly providing calibration and opacity tables for the LABOCA observations, 
as well as E. Dwek and R. Smith for enlightening discussion about X-ray radiated dust.
B.P. acknowledges the support by the Deutsche Akademie der Naturforscher Leopoldina (Halle, Germany) under grant BMBF-LPD 9901/8-170.\\
This publication is based on data acquired with the Atacama Pathfinder Experiment (APEX). 
APEX is a collaboration between the Max-Planck-Institut f\"{u}r Radioastronomie, the 
European Southern Observatory, and the Onsala Space Observatory.
This research has made use of SAOImage DS9, developed 
by SAO; 
and Astrophysics Data System Bibliographic Services by SAO/NASA.

\label{lastpage}

\end{document}